\documentclass[dvips]{acta}
\usepackage{supertabular,lscape}
\usepackage{amssymb}
\usepackage{amsmath}
\usepackage{epstopdf}
\usepackage{graphicx}

\SetPages{0}{0}

\SetVol{68}{2018}

\begin{document}

\begin{Titlepage}
\Title{V2676 Oph: Estimating physical parameters of a moderately fast nova}
%
%

\Author{Ashish Raj, M. Pavana, U. S. Kamath, G. C. Anupama}{Common Affiliation: Indian Institute of Astrophysics, II Block Koramangala, Bangalore 560034, India\\
e-mail: ashish.raj@iiap.res.in} and
\Author{F. M. Walter}
{Affiliation: Department of Physics and Astronomy, Stony Brook University, Stony Brook, NY 11794-3800, USA}

\Received{Month Day, Year}
\end{Titlepage}

\Abstract{Using our previously reported observations, we derive some physical parameters of the moderately fast nova V2676 Ophiuchi 2012 \#1. The best-fit CLOUDY model of the nebular spectrum 
obtained on 2015 May 8 shows a hot white dwarf source with T$_{BB}$ $\sim$ 1.0 $\times$ 10$^5$ K 
having a luminosity of 1.0 $\times$ 10$^{38}$ ergs s$^{-1}$. Our abundance analysis shows that the ejecta are significantly
enhanced relative to solar, He/H = 2.14, O/H = 2.37, S/H = 6.62 and Ar/H = 3.25. The ejecta mass is estimated to be 1.42 $\times$ 10$^{-5}$ M$_{\odot}$.
The nova showed a pronounced dust formation phase after 90 days from discovery. 
The J-H and H-K colors were very large as compared to other molecule- and dust-forming novae in recent years. The dust temperature and mass at two epochs have 
been estimated from spectral energy distribution (SED) fits to infrared photometry.

}
{optical: spectra - line : identification - stars : novae, cataclysmic variables - stars :
individual (V2676 Oph) - techniques : spectroscopic.}

\section{Introduction}
All novae are members of close binary systems containing white dwarfs (WDs) as the primary stars and late type main sequence (MS) stars or giants
as the secondaries or companions. A close binary may be detached, semi-detached, or a contact system depending on whether the companion star fills
its Roche lobe. 
The classical novae (CNe) are semi-detached binary systems with a late-type main sequence star (the secondary) transferring material
via Roche lobe overflow to a white-dwarf (the primary). The mass transfer results in the formation of an accretion disc around
the white dwarf. The runaway thermonuclear reactions on the white dwarf surface give rise to the cataclysmic outburst, the sudden brightening,
observed in these systems. Novae brighten by 10 - 12 magnitudes in a few hours and subsequently fade back to the original faint level over a period lasting 
several months to years. Evidences for early re-formation of the accretion disk, and presence of a circumbinary disk have been found in
recent times, (e.g. U Sco: Anupama et al. 2013 and references therein).

Novae provide a unique opportunity to study the interaction of changing radiation field with expanding ejecta spanning a wide
range of physical parameters like electron number density and temperature. There are different phases during the nova evolution viz. Fireball phase, when the ejecta is 
optically thick and behaves like a fireball which radiates as a hot blackbody of temperature 6000 - 10000 K ;
optically thin emission phase when the nova shell expands, becomes less dense and transparent. We may also see molecule and dust 
formation phase in some novae.
 It is generally accepted that the CNe binary system contains a carbon-oxygen
(CO) or oxygen-neon (ONe) WD primary and a low mass secondary.
Based on the optical spectra, novae are classified mainly in two categories Fe II and He/N (Williams 2012). But there are some hybrid novae which have been reported recently showing both the 
characteristic of Fe II and He/N novae in one outburst (see Raj et al. 2015a, Williams et al. 2017 and references therein). 

Nova V2676 Oph was discovered in outburst on 2012 March 25.789 UT at V = 12.1 (Nishimura et al. 2012). 
The optical spectra clearly indicate (from the numerous Fe II lines) that the nova belongs to Fe II spectral class. 
The distance to the nova was estimated to be between 6.9 and 7.9 kpc (Raj et al. 2017).
The nova showed molecule formation and subsequently dust formation during the outburst (see Kawakita et al. 2017 and references therein).
We have used the CLOUDY photoionization code, C13.04 (Ferland et al. 2013) for the abundance analysis in V2676 Oph.
This method has been used earlier to determine the elemental abundances for novae by modeling the observed
spectra e.g. LMC 1991 (Schwarz 2001), QU Vul (Schwarz et al. 2002), V1974 Cyg (Vanlandingham et al. 2005), V838 Her \& V4160 Sgr (Schwarz et al. 2007a), V1186 Sco
(Schwarz et al. 2007b), V1065 Cen (Helton et al. 2010) and RS Oph (Das \& Mondal 2015, Mondal et al. 2018).

\section{Data}

In this paper we have used the $JHK$ photometric and optical spectroscopic observations from Raj et al. (2017), and MIR photometry 
at 7.7, 8.8 and 12.4 $\mu$m reported by Kawakita et al. (2017). Photoioinization modeling of the nebular spectrum on 2015 May 8 (Raj et al. 2017) with a 
resolution of about 3\AA\ using the COSMOS long slit spectrograph at CTIO is done to estimate 
the physical parameters of the nova ejecta. This spectrum has been smoothed to reduce the noise. 
Spectral Energy Distribution (SED) plots are made to estimate the dust temperature.The outline of the paper is as follows: Sections 3 and 4
describe the results obtained from these observations and a summary and discussion is presented in
Section 5.

\section{CLOUDY modeling}
We have used the photoionization code CLOUDY, C13.04 (Ferland et al. 2013) to model the nebular phase spectrum of V2676 Oph taken on 2015 May 8, 1138 days after outburst.
We have chosen the nebular spectrum as there will be minimum dust emission at this stage which can otherwise contaminate the emission from other elements.
The modeled spectrum is compared with the observed spectrum and the elemental abundances, density, source luminosity, source 
temperature and the ejected mass are estimated. 

We define a set of parameters that specify the initial physical conditions of the source and the
ejected shell. These values are used by the photoionization code CLOUDY to generate a synthetic spectrum by solving the equations of thermal and statistical equilibrium from the non-LTE, 
illuminated gas clouds. These parameters include the shape and intensity of the external 
radiation field striking a cloud, the chemical composition of the gas, and the geometry of the gas, including its radial extent and the dependence of density on the radius. The density (hydrogen), 
radii, geometry, distance, covering factor, filling factor and elemental abundances define the physical conditions of the shell.  
The radial variation of hydrogen density, n(r) and filling factor, f(r) is given by:
\begin{align}
\label{eqn1}
\begin{split}
n(r) = n(r_{0})(\dfrac{r}{r_{0}})^{\alpha}\\
f(r) = f(r_{0})(\dfrac{r}{r_{0}})^{\beta}
\end{split}
\end{align}
where r$_{0}$ is the inner radius, $\alpha$ and $\beta$ are exponents of power laws. Here, $\alpha$ = -3  for a shell undergoing ballistic expansion, and the filling factor power 
law exponent, $\beta$ = 0 similar to previous studies (Schwarz 2002; Vanlandingham et al. 2005; Helton et al. 2010; Das \& Mondal 2015, Mondal et
al. 2018). 

We assume the geometry of the expanding shell to be spherically symmetric and illuminated by the central source with a 
total luminosity of about 10$^{38}$ erg s$^{-1}$ for the model. The free parameters, like hydrogen density, effective blackbody temperature and 
abundances of the elements seen in the observed spectrum e.g. He, O, S and Ar (remaining elements fixed at solar values) are varied to obtain the best-fit parameters. We assume the ejecta to comprise of two different density regions, the 
higher density to fit the low ionization lines, and the lower density to fit the high ionization lines, as the ejecta is non-homogeneous in density. 
The covering factor is set in both the regions such that the sum is always less than or equal to 1. Most of the parameters are kept constant in both the regions, so that the number of free 
parameters from both the regions could be reduced. The modeled line ratios are obtained by adding the line ratios of each region after multiplying by its covering factor. The inner and outer radii 
of the ejected shell are set by minimum and maximum expansion velocities obtained using the FWHM of the emission lines (400 to 2380 km s$^{-1}$) similar to the study by Helton et al. (2010), 
and the time (t = 1138 day). \\
\\
As CLOUDY uses many parameters which are interdependent to generate a spectrum, it is difficult to validate the final spectrum visually (Fig. 1). 
Thus, the best fit model is obtained by calculating $\chi^{2}$ and reduced $\chi^{2}$:
\begin{align}
\begin{split}
\label{eqn2}
\chi^{2} = \sum_{i=1}^{n} \dfrac{(M_{i} - O_{i})^{2}} {\sigma^{2}_{i}}\\
\chi^{2}_{red} = \dfrac{\chi^{2}}{\nu}
\end{split}
\end{align}
\\
M$_{i}$ \& O$_{i}$ -- modeled \& observed line ratios, respectively\\
$\sigma_{i}$ is the error in observed flux ratio\\
n -- number of observed lines\\
n$_{p}$ -- number of free parameters\\
$\nu$ = n - n$_{p}$, degrees of freedom.\\
The observed and modeled flux ratios with the $\chi^{2}$ values are given in Table 1. 

To match the predicted luminosities with the reddening-corrected observed flux, we assume a distance of $\sim$ 7 kpc for V2676 Oph. 
The measured line fluxes were dereddened for E(B - V ) = 0.93 (Raj et al. 2017) and compared with the output to calculate $\chi^{2}$ for the best fit. 
The errors associated with flux calibration for different wavelength regions are minimized from flux ratios of the modeled and observed 
prominent hydrogen lines, e.g., H$\alpha$ in the optical region. 
Using the following relation (e.g., Schwarz et al. 2001; Schwarz 2002; Das \& Mondal 2015):  
\begin{align}
\label{eqn3}
M_{shell} = n(r_{0})f(r_{0}) \int_{R_{in}}^{R_{out}} (\dfrac{r}{r_{0}})^{\alpha+\beta} 4 \pi r^{2} dr
\end{align}
the ejected mass is found to be 1.42$\times$10$^{-5}$ M$_{\odot}$.
The ejected mass is calculated for both the dense and diffuse regions, then multiplied by the corresponding covering factors and added to obtain the final value. 
The ejected mass estimated from 
this method is consistent with the values found by Raj et al. (2017). Using the relation below by Williams (1994),
\begin{align}
\label{eqn4}
\dfrac{F_{\lambda6300}}{F_{\lambda6364}} = \dfrac{(1 - e^{-\tau})}{(1 - e^{-\tau/3})}
\end{align}

we find that the optical depth $\tau$ of the ejecta for [O I]$\lambda$6300 
is 0.60 which is close to 0.49 obtained from CLOUDY. The electron temperature varies from 11237 K (first zone) to 4000 K (last zone) while the electron density is found to be of the order of 10$^{7}$ cm$^{-3}$ from the best-fit model.

\begin{table}[ht]
\caption{Observed and Best-fit CLOUDY Model line flux values relative to H$\alpha$. The $\chi^{2}$ values are calculated using equation 2.}
\label{table:1}
\begin{center}
\scriptsize
{\begin{tabular}{c  c  c  c c} 
 \hline
  \textbf{Line ID} & \boldmath{$\lambda$ ($\mu$m)} & \textbf{Observed} & \textbf{Modeled} & \boldmath{$\chi^{2}$} \\ [0.5ex] 
 \hline\hline
 $[$O I$]$ & 0.6300 & 2.70E$-$02 & 7.62E$-$02 & 2.51E$+$00\\ [0.25ex]
 $[$O I$]$ & 0.6364 & 1.00E$-$02 & 4.13E$-$02 & 1.25E$+$00\\ [0.25ex]
 H$\alpha$ & 0.6563 & 1.00E$+$00 & 1.00E$+$00 & 0.00E$+$00\\ [0.25ex]
 He I & 0.6678 & 3.06E$-$03 & 2.07E$-$02 & 6.99E$-$02\\ [0.25ex]
 $[$S II$]$ & 0.6716-0.6731 & 1.09E$-$02 & 1.20E$-$03 & 1.35E$-$02\\ [0.25ex]
 O I & 0.7002 & 5.46E$-$03 & 1.05E$-$02 & 2.06E$-$03\\ [0.25ex]
 He I & 0.7065 & 7.82E$-$03 & 1.50E$-$02 & 2.43E$-$02\\ [0.25ex]
 $[$Ar III$]$ & 0.7134 & 9.28E$-$03 & 3.32E$-$02 & 1.07E$-$01\\ [0.25ex]
 $[$Ar IV$]$ & 0.7237 & 6.94E$-$03 & 2.20E$-$02 & 1.12E$-$01\\ [0.25ex]
 He I & 0.7281 & 2.42E$-$03 & 7.01E$-$03 & 5.06E$-$03\\ [0.25ex]
 $[$O II$]$ & 0.7320-0.7330 & 4.07E$-$02 & 1.51E$-$01 & 3.69E$+$00\\ [0.25ex]
 $[$Ar III$]$ & 0.7751 & 1.18E$-$03 & 7.93E$-$03 & 1.24E$-$01\\ [0.25ex]
 H I (P10) & 0.9015 & 1.79E$-$03 & 1.93E$-$02 & 2.11E$-$01\\ [0.25ex]
 $[$S III$]$ & 0.9069 & 1.09E$-$02 & 1.16E$-$02 & 1.32E$-$04\\ [0.25ex]
 H I (P9) & 0.9229 & 4.51E$-$03 & 2.42E$-$02 & 6.03E$-$02\\ [0.25ex]
 $[$S III$]$ & 0.9531 & 2.96E$-$02 & 5.26E$-$02 & 1.45E$+$00\\ [0.25ex]
 H I (P7) & 1.0049 & 1.55E$-$03 & 5.87E$-$02 & 2.38E$-$01\\ [0.25ex]
 He II & 1.0124 & 1.01E$-$02 & 1.01E$-$03 & 1.44E$-$02\\ [0.25ex]
\hline
\end{tabular}}
\end{center}
\end{table}

\begin{table}[ht]
\caption{Best-fit CLOUDY Model parameters. The elemental abundance values are in logarithmic scale relative to hydrogen. Elements not listed here are assumed to have solar abundances. The number of lines used to determine abundances are given within brackets following the abundance value. }
\label{table:2}
\begin{center}
\scriptsize
{\begin{tabular}{c  c} 
 \hline
  \textbf{Parameter} & \textbf{Value}  \\ [0.5ex] 
 \hline\hline 
 T$_{BB}$ ($\times$ 10$^{5}$ K) & 1.00 \\ [0.25ex]
Source luminosity ($\times$ 10$^{38}$ erg/s) & 1.00 \\ [0.25ex]
Clump Hydrogen density ($\times$ 10$^{8}$ cm$^{-3}$) & 3.16 \\ [0.25ex]
Diffuse Hydrogen density ($\times$ 10$^{8}$ cm$^{-3}$) & 1.00 \\ [0.25ex]
Covering factor (clump) & 0.80 \\ [0.25ex]
Covering factor (diffuse) & 0.20 \\ [0.25ex]
$\alpha$ & -3.00 \\ [0.25ex]
Inner radius ($\times$ 10$^{15}$ cm) & 3.79 \\ [0.25ex]
Outer radius ($\times$ 10$^{16}$ cm) & 2.34 \\ [0.25ex]
Filling factor  & 0.05 \\ [0.25ex]
He/He$_{\odot}$ & 2.14 (4) \\ [0.25ex]
Ar/Ar$_{\odot}$ & 2.37 (3)\\ [0.25ex]
O/O$_{\odot}$ & 6.62 (4)\\ [0.25ex]
S/S$_{\odot}$ & 3.25 (3)\\ [0.25ex]
Ejected mass ($\times$ 10$^{-5}$ M$_{\odot}$) & 1.42 \\ [0.25ex]
Number of observed lines (n) & 18\\ [0.25ex]
Number of free parameters (n$_{p}$) & 9\\ [0.25ex]
Degrees of freedom ($\nu$) & 9\\ [0.25ex]
Total $\chi^{2}$ & 9.87\\ [0.25ex]
$\chi^{2}_{red}$ & 1.10\\ [0.25ex]
\hline
\end{tabular}}
\end{center}
\end{table}

The abundance values and other parameters obtained from the model are given in Table 2 in logarithm scale. 
The derived values indicate that the helium, argon, oxygen and sulphur abundances are all enhanced 
relative to solar. These abundance values are approximations to the true values because the calculation is based on three or four observed lines only. 
The abundance solutions are also sensitive to changing opacity and conditions of the ejecta. 
Though the model has low value of reduced $\chi^{2}$, it provides only a rough estimate of abundances. The Paschen lines are overestimated. Also, lack of multi-wavelength spectra results in possibly not-so-accurate abundance values. However, the model gives estimates of several parameters including temperature, luminosity, ejected mass, opacity (see Table 1 and 2) which are in the range of values expected from Fe-II novae. 
\section{J-H and H-K colors, dust temperature and mass}
The optical and near-IR spectrophotometric evolution of nova V2676 Oph presented in Raj et al. (2017) clearly showed the dust formation about 90 days since outburst (Fig 1. Raj et al. 2017).
We study here the IR color evolution to understand the nature of the dust. Fig. 2 shows the J-H and H-K colors spanning a period of $\sim$ 500 days.
The J-H and H-K colors reached a maximum value of $\sim$ 4.5 mag at 155 days and $\sim$ 3.8 mag at 
170 days, respectively, from the outburst. The observed NIR excess suggests that a large amount of optically thick dust formed in the nova ejecta.
We compare these values with the other molecule and dust forming novae in recent times and find that these may be the largest J-H and H-K values observed so far (see Table 3). 
After reaching a peak, the J-H and H-K colors started to show a downward trend, which indicate destruction of the dust molecules due to radiation from the central source. Another nova, V1324 Sco has shown gamma-ray emission as well as dust formation (but no molecule formation) had similar colors, J-H = 4.9 and H-K = 3.4 (Finzell et al. 2018) 
as seen in V2676 Oph.

\begin{table*}[ht]
\caption{The maximum values of J-H and H-K colors for recent molecule- and dust-forming novae.}
\label{table:1}
\begin{center}
\scriptsize
{\begin{tabular}{c  c  c  c c} 
 \hline
   & V2615 Oph & V496 Sct & V5584 Sgr & V2676 Oph \\ [0.5ex] 
 \hline
 J-H & 1.7 & 1.5 & 0.3 & 4.5\\ [0.25ex]
 H-K & 2.3 & 2.3 & 0.4 & 3.8\\ [0.25ex]
 Spectral Class & Fe II & Fe II & Fe II & Fe II\\ [0.25ex]
 Grain type & carbon & carbon & carbon  & carbon and silicate\\ [0.25ex]
 Dust Mass & -- & 1-5 $\times$ 10$^{-10}$ M$_\odot$  & 7.35 $\times$ 10$^{-9}$ M$_\odot$ & 2.1-2.7 $\times$ 10$^{-8}$ M$_\odot$\\ [0.25ex]
 References& Das et al. (2009) & Raj et al. (2012) & Raj et al. (2015b) & Kawakita et al. (2017)\\ [0.25ex]
 
\hline
\end{tabular}}
\end{center}
\end{table*}

A sharp fall in the optical light curve about 90 days after the outburst clearly indicate the onset of the dust formation (Raj et al. 2017). We expect to see a rise in the JHK band magnitudes during dust 
formation (sharp fall in the optical), but the JH bands show 
smaller drop and a flattening in K band after 90 days. This indicate the self-absorption in the JHK bands i.e. varying extinction in these bands so we can not use the JHK magnitudes to 
construct the SED for estimating the dust temperature and mass during this time period. Therefore, we use the observed HK magnitudes on 2013 June 20 (day 452), K magnitude on 2014 May 16 
(day 782) and the IR photometric observations from Kawakita et al. (2017), 
to construct the SED of the dust component in the ejecta (see Fig. 3). The thermal emission from the dust is seen increasing up to the K band and it peaks at
even longer wavelengths. We estimate the temperature of the dust shell on these two epochs as
700 $\pm$ 100 K and 400 $\pm$ 100 K, respectively. However, the estimate of temperature for the dust shell
may have a large uncertainty as we have used only five and four wavelengths, respectively, to fit the SED and the dust might be
contributing at larger wavelengths beyond 12.4 $\mu$m.
The mass of the dust shell is calculated from the thermal component of the SED of 2013 June 20 shown in Fig. 3 (upper panel). 
Here we assume that the dust is mainly made of carbon particles (amorphous carbon has higher condensation temperature than astronomical silicates) having size
less than 1 $\mu$m with a density of 2.25 gm cm$^{-3}$.
The expression given for the mass of the dust shell in Woodward et al. (1993) and the distance range 6.9 - 7.9 kpc (Raj et al. 2017) have been used to estimate the dust
mass as M$_{dust}$ = (2.1-2.7) $\times$ 10$^{-8}$ M$_\odot$.   This shows 
that a large amount of optically thick dust was formed between June 2012 to December 2012 in the nova ejecta.  Note that we have estimated the mass of the amorphous carbon only. This type of dust generally forms earlier when the ejecta temperature is high. Kawakita et al. (2017) consider various combinations of carbon and silicate dust. 

\begin{figure}[htb]
\includegraphics[width=1.0\columnwidth]{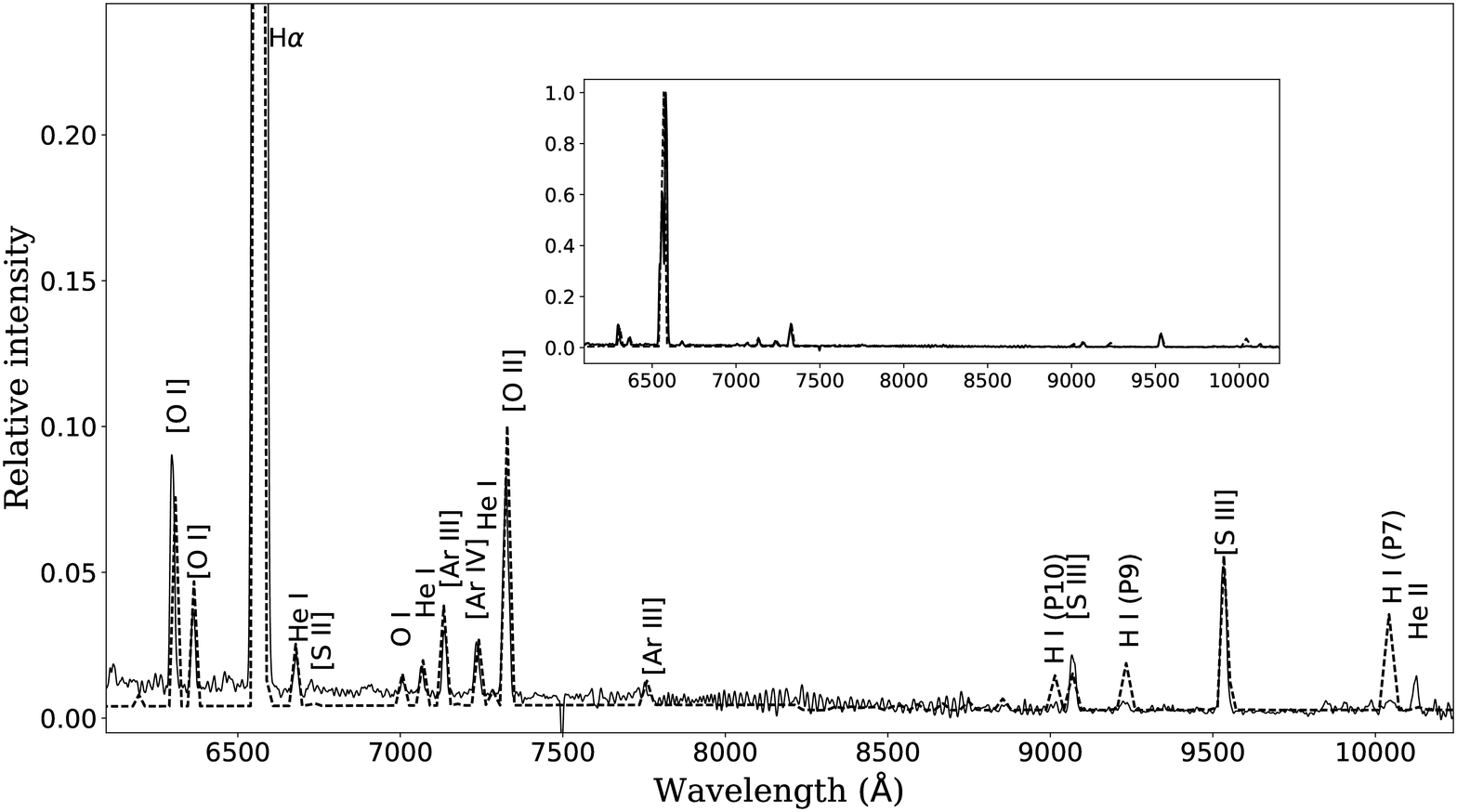}
\FigCap{The CLOUDY model (dotted line) fit to the observed optical spectra (solid lines) of V2676 Oph
observed on 2015 May 8. The Paschen lines are overestimated. The spectra are normalized to H$\alpha$
(see section 3 for details). }\label{figOne}
\end{figure}

\begin{figure}[htb]
\includegraphics{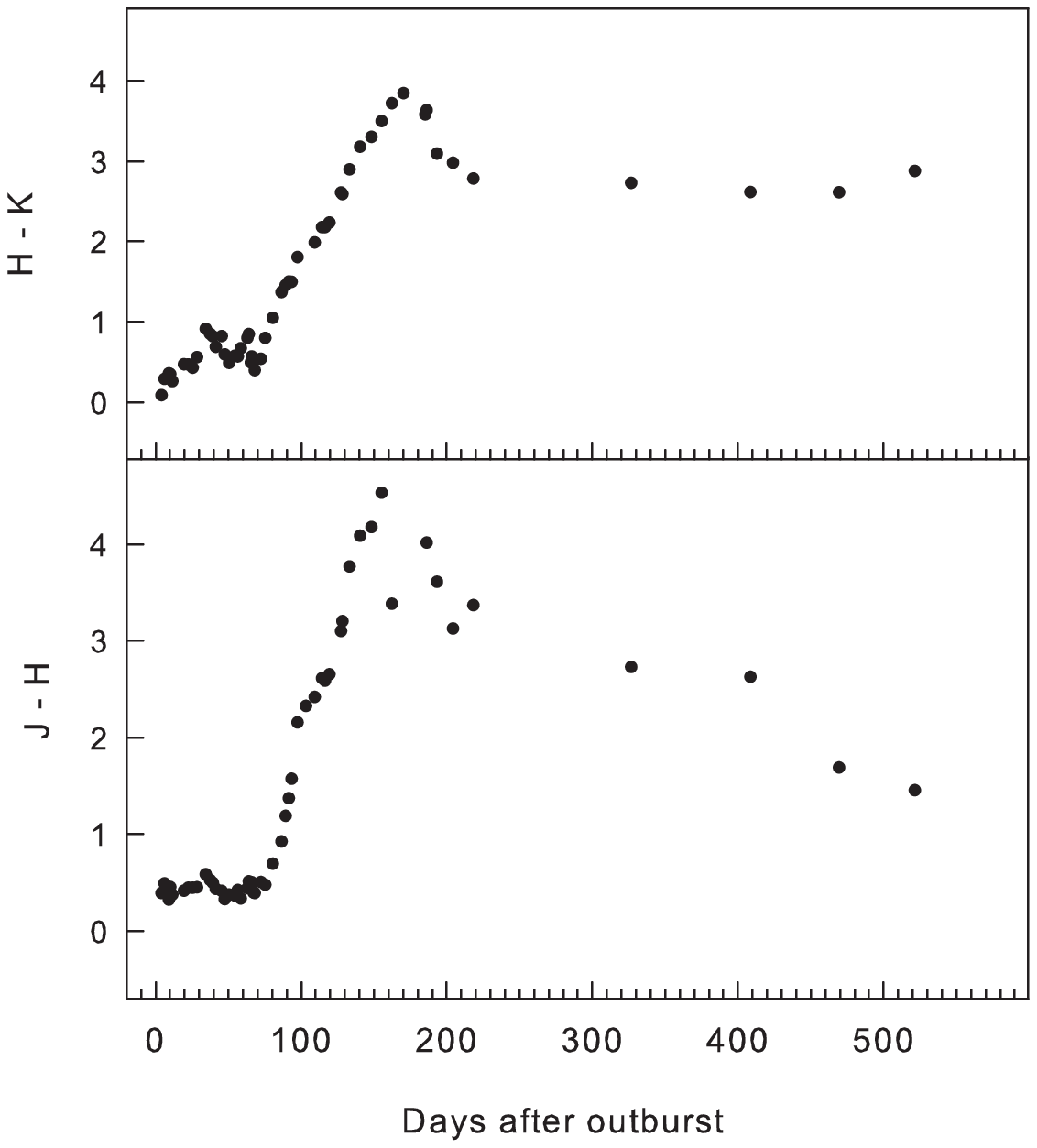}
\FigCap{The J-H and H-K colors of V2676 Oph are shown based on the data from Raj et al. (2017).}
\end{figure}

\begin{figure}[htb]
\includegraphics{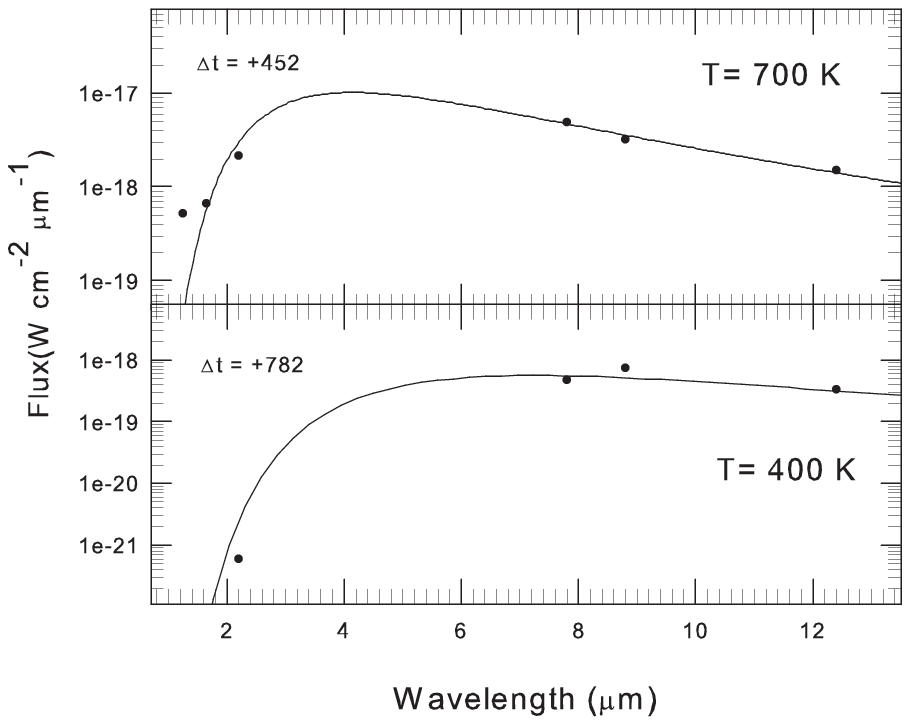}
\FigCap{The SED for nova V2676 Oph is shown for 2013 June 20
(upper panel) and 2014 May 16 (lower panel). The $JHK$ data are taken from Raj et al. (2017) and the flux values at 7.7, 8.8 and 12.4 $\mu$m are taken from Kawakita et al. (2017).}
\end{figure}

\section{Summary and Discussion}

Using the CLOUDY code, physical parameters e.g. optical depth, electron temperature, mass of the ejecta are estimated for V2676 Oph.
Assuming a spherical geometry of the ejecta consisting two different shells of different densities a set of spectra are generated by varying several parameters. 
We compared the model-generated spectra with observed spectra to estimate the best fit parameters using $\chi^{2}$ technique. 
The WD temperature was estimated as 1.0 $\times$ 10$^{5}$ K having luminosity of 1.0 $\times$ 10$^{38}$ ergs s$^{-1}$.
The ejecta are significantly enhanced, relative to solar, in
helium, argon, oxygen and sulphur. We estimate an ejected mass as 1.42 $\times$ 10$^{-5}$ M$_{\odot}$ which suggests that a low mass WD is
associated with the nova V2676 Oph. 
To understand the thermonuclear runaway (TNR) process which is responsible for the nova explosion, it is very important to have a complete knowledge of elemental abundances in the
ejecta, the composition of accreted material on the WD as we can't rule out the possibility
of mixing the ejecta with the WD material.

The J-H and H-K colors indicate a large reddening due to the dust formed in the nova ejecta between June 2012 to December 2012. 
Kawakita et al. (2017) reported carbon and oxygen-rich dust grains in the nova ejecta and they also pointed out that hosting white-dwarf (WD) for V2676 Oph is CO-rich 
as no [Ne II] line at 
12.8 $\mu$m is detected in later stage. This is further supported by the slow evolution of the light curve and large ejecta mass estimated in this paper. 
Typically the CO WDs have lower mass compared to ONe WDs but the observed isotopic ratios of carbon and nitrogen in 
V2676 Oph indicates large WD mass $\sim$1.15 M$_{\odot}$ for V2676 Oph 
(Kawakita et al. 2017 and references therein). So, if one looks at the traditional parameters e.g. light curve, ejecta mass, copious dust production then it seems that the WD in V2676 Oph is a
low mass CO 
WD. However, based on the observed isotopic ratio of C and N, coupled with the latest TNR models (Jose \& Hernanz 1998, 2007; Jose et al. 2006; Denissenkov et al. 2014 ), 
it seems that the WD could be as massive as $\sim$1.15 M$_{\odot}$. This stresses the 
importance of developing better TNR models for the nova outburst. Similarly, the method of relying on just the traditional parameters for determining the nature of the WD may need some improvement.

\end{document}